\documentclass{ifacconf}
\usepackage{graphicx}      % include this line if your document contains figures
\usepackage{natbib}        % required for bibliography
\usepackage{amssymb,amsmath}
\usepackage{mathdots}
\usepackage{mathtools}
\usepackage{amsfonts,commath}
\usepackage{microtype}
\usepackage[T1]{fontenc}
\usepackage{booktabs}
\usepackage{enumerate}

\usepackage{tikz}
\usepackage{pgfplots}
\pgfplotsset{every axis/.append style={line width=1pt}}
\usetikzlibrary{shapes,shadows,arrows,backgrounds,patterns,positioning}

\usepackage[abs]{overpic}

\usepackage[applemac]{inputenc}

\newcommand\inM[1]{#1\in\mathcal{M}}
\newcommand\inN[2]{#1\in\mathcal{N}_{#2}}

\newcommand\mbb[1]{\mathbb{#1}}

\newcommand{\ts}[1]{{\textnormal{#1}}}
\newcommand{\ie}{\emph{i.e.},}

\newcommand{\mb}{\mathbf}
\newcommand{\mc}{\mathcal}

\DeclarePairedDelimiter{\interiorpars}{(}{)}
\newcommand{\interior}{\operatorname{interior}\interiorpars}

\newlength{\hatchspread}
\newlength{\hatchthickness}
\newlength{\hatchshift}
\newcommand{\hatchcolor}{}
\tikzset{hatchspread/.code={\setlength{\hatchspread}{#1}},
         hatchthickness/.code={\setlength{\hatchthickness}{#1}},
         hatchshift/.code={\setlength{\hatchshift}{#1}},% must be >= 0
         hatchcolor/.code={\renewcommand{\hatchcolor}{#1}}}
\tikzset{hatchspread=3pt,
         hatchthickness=0.4pt,
         hatchshift=0pt,% must be >= 0
         hatchcolor=black}
\pgfdeclarepatternformonly[\hatchspread,\hatchthickness,\hatchshift,\hatchcolor]% variables
   {nwl}% name
   {\pgfqpoint{\dimexpr-2\hatchthickness}{\dimexpr-2\hatchthickness}}% lower left corner
   {\pgfqpoint{\dimexpr\hatchspread+2\hatchthickness}{\dimexpr\hatchspread+2\hatchthickness}}% upper right corner
   {\pgfqpoint{\dimexpr\hatchspread}{\dimexpr\hatchspread}}% tile size
   {% shape description
    \pgfsetlinewidth{\hatchthickness}
    \pgfpathmoveto{\pgfqpoint{0pt}{\dimexpr\hatchspread+\hatchshift}}
    \pgfpathlineto{\pgfqpoint{\dimexpr\hatchspread+0.15pt+\hatchshift}{-0.15pt}}
    \ifdim \hatchshift > 0pt
      \pgfpathmoveto{\pgfqpoint{0pt}{\hatchshift}}
      \pgfpathlineto{\pgfqpoint{\dimexpr0.15pt+\hatchshift}{-0.15pt}}
    \fi
    \pgfsetstrokecolor{\hatchcolor}
    \pgfusepath{stroke}
   }
\pgfdeclarepatternformonly[\hatchspread,\hatchthickness,\hatchshift,\hatchcolor]% variables
   {nwld}% name
   {\pgfqpoint{\dimexpr-2\hatchthickness}{\dimexpr-2\hatchthickness}}% lower left corner
   {\pgfqpoint{\dimexpr\hatchspread+2\hatchthickness}{\dimexpr\hatchspread+2\hatchthickness}}% upper right corner
   {\pgfqpoint{\dimexpr\hatchspread}{\dimexpr\hatchspread}}% tile size
   {% shape description
    \pgfsetlinewidth{\hatchthickness}
    \pgfpathmoveto{\pgfqpoint{0pt}{\dimexpr\hatchspread+\hatchshift}}
    \pgfpathlineto{\pgfqpoint{\dimexpr\hatchspread+0.15pt+\hatchshift}{-0.15pt}}
    \ifdim \hatchshift > 0pt
      \pgfpathmoveto{\pgfqpoint{0pt}{\hatchshift}}
      \pgfpathlineto{\pgfqpoint{\dimexpr0.15pt+\hatchshift}{-0.15pt}}
    \fi
    \pgfsetstrokecolor{\hatchcolor}
    \pgfsetdash{{1pt}{1pt}}{0pt}% dashing cannot work correctly in all situation this way
    \pgfusepath{stroke}
   }
\pgfdeclarepatternformonly[\hatchspread,\hatchthickness,\hatchshift,\hatchcolor]% variables
   {nel}% name
   {\pgfqpoint{\dimexpr-2\hatchthickness}{\dimexpr-2\hatchthickness}}% lower left corner
   {\pgfqpoint{\dimexpr\hatchspread+2\hatchthickness}{\dimexpr\hatchspread+2\hatchthickness}}% upper right corner
   {\pgfqpoint{\dimexpr\hatchspread}{\dimexpr\hatchspread}}% tile size
   {% shape description
    \pgfsetlinewidth{\hatchthickness}
    \pgfpathmoveto{\pgfqpoint{\dimexpr\hatchshift-0.15pt}{-0.15pt}}
    \pgfpathlineto{\pgfqpoint{\dimexpr\hatchspread+0.15pt}{\dimexpr\hatchspread-\hatchshift+0.15pt}}
    \ifdim \hatchshift > 0pt
      \pgfpathmoveto{\pgfqpoint{-0.15pt}{\dimexpr\hatchspread-\hatchshift-0.15pt}}
      \pgfpathlineto{\pgfqpoint{\dimexpr\hatchshift+0.15pt}{\dimexpr\hatchspread+0.15pt}}
    \fi
    \pgfsetstrokecolor{\hatchcolor}
    \pgfusepath{stroke}
   }
\pgfdeclarepatternformonly[\hatchspread,\hatchthickness,\hatchshift,\hatchcolor]% variables
   {neld}% name
   {\pgfqpoint{\dimexpr-2\hatchthickness}{\dimexpr-2\hatchthickness}}% lower left corner
   {\pgfqpoint{\dimexpr\hatchspread+2\hatchthickness}{\dimexpr\hatchspread+2\hatchthickness}}% upper right corner
   {\pgfqpoint{\dimexpr\hatchspread}{\dimexpr\hatchspread}}% tile size
   {% shape description
    \pgfsetlinewidth{\hatchthickness}
    \pgfpathmoveto{\pgfqpoint{\dimexpr\hatchshift-0.15pt}{-0.15pt}}
    \pgfpathlineto{\pgfqpoint{\dimexpr\hatchspread+0.15pt}{\dimexpr\hatchspread-\hatchshift+0.15pt}}
    \ifdim \hatchshift > 0pt
      \pgfpathmoveto{\pgfqpoint{-0.15pt}{\dimexpr\hatchspread-\hatchshift-0.15pt}}
      \pgfpathlineto{\pgfqpoint{\dimexpr\hatchshift+0.15pt}{\dimexpr\hatchspread+0.15pt}}
    \fi
    \pgfsetstrokecolor{\hatchcolor}
    \pgfsetdash{{1pt}{1pt}}{0pt}% dashing cannot work correctly in all situation this way
    \pgfusepath{stroke}
   }
\newif\ifproves
\provestrue

\begin{document}
\begin{frontmatter}

\title{Nested distributed MPC\thanksref{footnoteinfo}}

\thanks[footnoteinfo]{Work supported by the Harry Nicholson PhD Scholarship, Department of Automatic Control \& Systems Engineering,
University of Sheffield, and Doctoral Scholarship from CONICYT--PFCHA/ Concurso para Beca de Doctorado en el Extranjero--72150125.}

\author[First]{Pablo R. Baldivieso Monasterios}
\author[First]{Bernardo Hernandez}
\author[First]{Paul A. Trodden} 

\address[First]{Department of Automatic Control \& Systems Engineering,\\ University of Sheffield, Sheffield S1 3JD, UK \\(e-mail \{prbaldivieso1, bahernandezvicente1, p.trodden\}@sheffield.ac.uk)}

\begin{abstract}                % Abstract of not more than 250 words.
We propose a distributed model predictive control  approach for linear
time-invariant systems coupled via dynamics. The proposed approach
uses the tube MPC concept for robustness to handle the disturbances
induced by mutual interactions between subsystems; however, the main
novelty here is to replace the conventional linear disturbance
rejection controller with a second MPC controller, as is done in
tube-based nonlinear MPC. In the distributed setting, this has the
advantages that the disturbance rejection controller is able to
consider the plans of neighbours, and the reliance on explicit robust
invariant sets is removed.
\end{abstract}


\end{frontmatter}
%===============================================================================

\section{Introduction}
Model Predictive Control (MPC) is a mature and popular control
technique~\citep{RM_mpc_book, Mayne14} that excels in situations where it is
prohibitively difficult to design a control law off-line: for example,
in the presence of constraints. MPC is inherently, however,
a \emph{centralized} control technique, and so its applicability to
large-scale systems is limited by the fact that the
controller would have to model, sense and control the whole
plant. For this reason, significant attention has been given to
non-centralized MPC, including decentralized,
distributed and hierarchical forms~\citep{Scattolini09}. The main challenge is how to coordinate the control actions of independent MPC-based
controllers, in order that
the overall system is stable and satisfies constraints. Many proposals
have been made (see~\citet{Scattolini09,CSM+13} for excellent surveys), and
broadly differ according to the nature or source of the coupling
between subsystems and the algorithmic approach taken to coordinate
control actions~\citep{MN14}.

The problem tackled in this paper is the fundamental one of
controlling dynamically coupled linear time-invariant systems. The
problem is challenging because the states and inputs of one subsystem
affect others too; therefore, the straightforward application of MPC,
even with terminal conditions~\citep{RM_mpc_book}, does not guarantee
constraint satisfaction and stability. A popular approach
is to decompose and distribute the MPC problem (or its dual)
among the different controllers, and solve the problem iteratively at
each time step---with information exchange between controllers---until
feasibility or optimality is obtained~\citep{MN14}; however, the
price to pay is large amounts of communication, slow convergence (of
iterates) in large systems, and a long time to solve to MPC problem at
each step.

In pursuit of iteration-free methods that still achieve desirable
guarantees, a few authors~\citep{FS12,RF12,TBM16,HVT16} have exploited ideas
from \emph{robust} MPC, and particularly tube-based
MPC~\citep{MSR05}. The basic idea is---considering the mutual
interactions as exogenous disturbances---to augment the conventional
MPC control law with an ancillary, disturbance rejection term, computed off-line
and based on the theory of disturbance-invariant sets. The main
drawback is the conservatism induced by taking a robust approach to
what is a nominal problem, and research efforts have focused on ways
in which to reduce this and improve performance: \citet{FS12} employ
reference trajectories, and consider the disturbances as deviations
from these. \citet{RF12} employ the tube concept \emph{twice},
designing two disturbance rejection controllers: the first to minimize
deviations between a planned nominal trajectory and planned perturbed
trajectory, and the second to minimize deviations between the latter
and the true perturbed trajectory. \citet{TBM16} propose a more
straightforward design, with only one disturbance rejection controller
and no reference trajectories, but optimize disturbance-invariant
sets on-line in order to reduce conservatism.

In this paper, we offer a new contribution to the family of tube-based
distributed MPC (DMPC) approaches. The main development here is to
replace the ancillary disturbance rejection controller---which is
linear in each of~\citet{FS12,RF12,TBM16}---with an ancillary MPC
controller, which operates in a nested fashion with the main
controller. This development is inspired by the approach
of~\citet{MKW+11} for tube-based \emph{nonlinear} MPC, which introduced
the idea of an ancillary MPC controller; in that approach, the
controller is needed because of the non-linearity of the system. Here
we employ the second controller for a different purpose, which also
leads to two advantages with respect to existing tube-based DMPC: the
ancillary controller is able to consider the plans of neighbouring
subsystems when optimizing the disturbance rejection control action;
perhaps more significantly, the need to explicitly compute and employ
disturbance-invariant sets---which are prohibitively complex objects
for high-dimension subsystems---is removed.
	
The problem statement is defined in
Section~\ref{sec:Preliminaries}. In Section~\ref{sec:control_alg}, the
nested DMPC approach is developed, including optimal control problems
and the distributed algorithm. Recursive feasibility and stability are
established in section~\ref{sec:stab_feas}. A comprehensive off-line
design method to select controller parameters is given in
Section~\ref{sec:constant_selec}, before an illustrative example of the approach is presented in Section~\ref{sec:sims}.

\emph{Notation:}  The sets of non-negative and positive reals are denoted, respectively, $\mbb{R}_{0}^+$ and $\mbb{R}^+$. $AX$ denotes the image of a set $X \subset \mbb{R}^n$ under the linear mapping $A:\mbb{R}^n \mapsto \mbb{R}^p$, and is given by $\{ Ax : x \in X\}$.For $X, Y \subset \mbb{R}^n$, the Minkowski sum is $X \oplus Y\triangleq\{ x + y: x \in X, y \in Y \}$; for $Y \subset X$. For $X \subset \mbb{R}^n$ and $a \in \mbb{R}^n$, $X\oplus a$ means $X \oplus \{a\}$. A polyhedron is an intersection of a finite number of half-spaces, and a polytope is a closed and bounded polyhedron.  A \emph{C-set} is a compact and convex set that contains the origin; in a \emph{PC-set}, the origin is within the interior. The C-set $L$ is said to be a summand of $K$ if there exists a set $M$ such that $K = L\oplus M$. A sequence is defined as $\mb{x} =\{x(0),x(1),\ldots\}$, the cardinality of which will be clear from the context. The notation $x_{-i}$ indicates a sequence without the $i$th member.
\section{Problem Statement}\label{sec:Preliminaries}

We consider the discrete-time
dynamics
\begin{equation}
  x^+=Ax+Bu
\label{eqn:large_scale}
\end{equation}
where $x \in \mathbb{R}^n$ and $u \in \mathbb{R}^m$ are the state and
control input, and $x^+$ is the state at the next time instance. This
system is partitioned or decomposed into $M$ non-overlapping
subsystems, in the sense that the state and input may be written $x =
(x_1,\ldots,x_M)$ and $u = (u_1,\ldots,u_M)$, where $x_i \in
\mathbb{R}^{n_i}$ and $u_i \in \mathbb{R}^{m_i}$ are the state and
input of subsystem $i$, $\sum_{\inM{i}}n_i =n$ and $\sum_{\inM{i}}m_i
=m$, and the dynamics of subsystem $i \in \mc{M}\triangleq
\{1,\dots,M\}$ may be written as
\begin{equation}
x_i^+=A_{ii}x_i+B_{ii}u_i+\sum_{j \neq i} A_{ij}x_j+B_{ij}u_j.
\label{eqn:agent_partition}
\end{equation}
In this equation, $A_{ij} \in \mathbb{R}^{n_i \times n_j}$, $B_{ij}
\in \mathbb{R}^{n_i \times m_j}$ are the relevant block elements of
$A$ and $B$. The summation term represents the interaction of the
states and inputs of other subsystems $(j\neq i)$ on the dynamics of
subsystem $i$; without loss of generality, the summation may be
performed over $j \in \mc{N}_i$, where $\mathcal{N}_i \triangleq \{ j
\in \mc{M} : [A_{ij}\ B_{ij}]\neq 0, j \neq i\}$ is the set of
\emph{neighbours} of $i$.

\begin{assum}
For each $i \in \mathcal{M}$, $(A_{ii},B_{ii}) $ is controllable.
\label{assump:contr}
\end{assum}

Each subsystem $i \in \mathcal{M}$ is subject
to local constraints on its states and inputs
\begin{align}\label{eqn:local_cons}
	x_i \in \mathbb{X}_i, && u_i \in \mathbb{U}_i.
\end{align}

\begin{assum}\label{assump:constraint_sets}
For each $i \in \mathcal{M}$, $\mathbb{X}_i$ and $\mathbb{U}_i$ are PC-sets.
\end{assum}

The control objective is to steer the states of all subsystems to the
origin while satisfying the constraints and minimizing the
infinite-horizon cost
\begin{equation}
	\sum_{k=0}^\infty\sum_{i\in\mathcal{M}} \ell_i(x_i(k),u_i(k))
	\label{eqn:iinfine_cost}
\end{equation}
where $\ell_i(x_i,u_i) \triangleq (1/2)(x_i^\top Q_i x_i + u_i^\top
R_i u_i)$ and $Q_i$, $R_i$ are positive definite for all $i \in
\mc{M}$.

\section{Nested Distributed MPC}\label{sec:control_alg}

The main challenge with respect to controlling the
system~\eqref{eqn:large_scale} with independent, decentralized
controllers is how to deal with the interactions, for the states and
inputs of one subsystem are affected by, and affect, others in the
system. The most direct approach ignores these interactions, and
employs the nominal prediction model
\begin{equation}
  \bar{x}^+_i = A_{ii}\bar{x}_i + B_{ii} \bar{u}_i
  \label{eq:nom_sys}
  \end{equation}
within an MPC optimization to provide the receding horizon control law
$u_i = \bar{\kappa}_i(x_i)$, obtained by applying the first control
$\bar{u}_i^0(0;x_i)$ in the optimized sequence. Ignoring interactions
in this way, however, can lead to constraint violations and even
instability, unless further actions or design steps are taken to
coordinate the actions of controllers~\citep{Scattolini09}.

An alternative approach is to treat all interactions as disturbances
to be rejected. The dynamic coupling between subsystems---arising
from the decomposition of the large-scale system---induces mutual
disturbances upon each subsystem; in fact, we may
re-write~\eqref{eqn:agent_partition} as the uncertain dynamics
\begin{equation}
  x_i^+=A_{ii}x_i+B_{ii}u_i+ w_i
  \label{eq:unc_sys}
\end{equation}
where $w_i \triangleq \sum_{j \in \mc{N}_i} (A_{ij}x_j +
B_{ij}u_j)$. This disturbance is, in view of the constraints on each
$x_j$ and $u_j$, contained within the set
\begin{equation}
  \label{eqn:disturbance_sets}
	\mbb{W}_i\triangleq \bigoplus_{j\in\mc{N}_i}A_{ij}\mbb{X}_j\oplus B_{ij}\mbb{U}_j,
\end{equation}
which, because of Assumption~\ref{assump:constraint_sets}, is bounded
and at least a C-set. The local control problem is then to regulate
the uncertain, constrained LTI system~\eqref{eq:unc_sys} which is
subject to bounded additive disturbances, and the direct application
of a robust MPC technique will (under suitable further assumptions)
lead to guaranteed feasibility and stability. For example, one could
employ the tube-based approach to robust MPC~\citep{MSR05}, which
retains the nominal model for predictions within an MPC problem with
restricted constraints (see the problem $\mbb{P}_i(\bar{x}_i)$ in the
next subsection), but augments the implicit control law with a linear,
disturbance rejection control law:
\begin{equation*}
u_i = \bar{\kappa}_i(\bar{x}_i) + K_i (x_i - \bar{x}_i).
  \end{equation*}
The latter term corrects for the errors introduced by neglecting the
disturbance (the interactions) in the predictions. The price to pay is
conservatism, for the controllers are designed to be robust to the
\emph{whole} space of possible states and inputs of other subsystems: neither the nominal MPC control law nor the linear disturbance
rejection controller take into account the planned states and/or
inputs of interacting subsystems. Hence, approaches that utilize
tubes~\citep{FS12,RF12,TBM16} have focused on ways in which the
conservatism can be reduced.

In this paper, we present a third way to this problem, with the aim of
retaining the desirable guarantees that a robust approach brings, but
lessening the conservatism and other drawbacks associated with this. In
particular, we propose a control law of the form
\begin{equation}
u_i = \kappa_i(x_i) = \bar{\kappa}_i(\bar{x}_i) + \hat{\kappa}_i (x_i - \bar{x}_i;
\bar{\mb{x}}_{-i}, \bar{\mb{u}}_{-i}),
\label{eq:law}%
  \end{equation}
which, inspired by~\citet{MKW+11}, replaces the linear disturbance
rejection control law of tube MPC with a \emph{second} predictive
control law. The second term still acts on the error $x_i - \bar{x}_i$
between the true (perturbed) state and the predicted (nominal) state,
but takes into account information shared by other subsystems about
their predicted states and inputs. These shared predictions are the
outputs of the \emph{first} predictive controller; hence, the
controllers for a subsystem work in a nested fashion.

The remainder of this section presents the approach, including the
optimal control problems and the algorithm. First, we require the
following assumption about the disturbance set, which is common in
tube-based MPC, but here effectively limits the strength of coupling
between subsystems:
\begin{assum}
For each $i \in \mc{M}$, $\mbb{W}_i\subset\interior{\mbb{X}_i}$.
\label{assump:disturbance}	
\end{assum} 

\subsection{Main optimal control problem}

The main optimal control problem for subsystem $i$ employs the nominal
model~\eqref{eq:nom_sys} to determine, in the presence of tightened
constraints, a nominal optimal control sequence and associated nominal
state predictions. Formally, this problem is $\mbb{P}_i(\bar{x}_i)$,
defined as
\begin{equation}
\bar{V_i}^0(\bar{x}_i) = \min_{\bar{\mb{u}}_i} \sum_{k =
  0}^{N-1}\ell_i (\bar{x}_i(k),\bar{u}_i(k) )
\label{eqn:mpc_outer}
\end{equation}
% V^f_i(\bar{x}_i(N)) +
subject to
\begin{subequations}
\begin{align}
  \bar{x}_i(0) &=\bar{x}_i, \label{eqn:cons_init}\\
  \bar{x}_i(k+1) &= A_{ii}\bar{x}_i(k)+B_{ii}\bar{u}_i(k), k = 0,\dots,N-1,\label{eqn:cons_dyn}\\
  \bar{x}_i(k) &\in \alpha_i^x \mbb{X}_i, k = 1,\dots,N-1,\label{eqn:cons_x}\\ 
  \bar{u}_i(k) &\in \alpha_i^u \mbb{U}_i, k = 1,\dots,N-1,\label{eqn:cons_u}\\
%  \bar{x}_i(N) &\in \mbb{X}_i^f. \label{eqn:cons_term}
  \bar{x}_i(N) &= 0. \label{eqn:cons_term}
\end{align}
\end{subequations}
In this problem, the decision variable $\mb{u}_i$ is the sequence of
(nominal) controls
\begin{equation*}
\bar{\mb{u}}_i = \{\bar{u}_i(0),\bar{u}_i(1),\dots,\bar{u}_i(N-1)\}.
\end{equation*}
The original state and input constraint sets, $\mbb{X}_i$ and
$\mbb{U}_i$, are scaled by factors $\alpha_i^x \in (0,1]$ and
  $\alpha_i^u \in (0,1]$ respectively, in order to preserve constraint
    satisfaction despite the neglecting of the disturbance
    (interaction) in the predictions. A detailed and comprehensive
    design procedure for these scalars is given in
    Section~\ref{sec:constant_selec}.
    
\begin{rem}
	For simplicity, we use the origin as terminal set; less restrictive conditions are subject to current research.
\end{rem}

The solution of $\mbb{P}_i(\bar{x}_i)$ at nominal state $\bar{x}_i$
yields the optimal control and state sequences
$\bar{\mb{u}}_i^0(\bar{x}_i) =
\{\bar{u}^0_i(0;\bar{x}_i),\ldots,\bar{u}^0_i(N-1;\bar{x}_i)\}$ and
$\bar{\mb{x}}_i^0(\bar{x}_i) =
\{\bar{x}^0_i(0;\bar{x}_i),\ldots,\bar{x}^0_i(N;\bar{x}_i)\}$. It also
defines the implicit control law
\begin{equation*}
\bar{\kappa}_i(\bar{x}_i) = \bar{u}^0_i(0;\bar{x}_i).
  \end{equation*}

In the next section, we define the ancillary optimal control problem
that yields the second part of the control law~\eqref{eq:law}.

\subsection{Ancillary optimal control problem}
\label{sec:anc_optimal_control}

The aim of the ancillary MPC controller is to reduce the error between
true states and predictions. This error is $e_i
\triangleq x_i - \bar{x}_i$, and evolves as
\begin{equation*}
{e}_i^+=A_{ii}{e}_i+B_{ii}f_i + \sum_{\inN{j}{i}} A_{ij}x_j+B_{ij}u_j
  \end{equation*}
where $f_i\triangleq u_i - \bar{u}_i$. In a conventional single tube
MPC controller approach, $f_i= K_i e_i$, but here we wish to replace
this simple linear controller with a controller that can account for
predictions of neighbouring subsystems. The above error dynamics are,
however, not suitable for use as a prediction model because of the
dependency on true states and inputs, $x_j$ and $u_j$, rather than shared
predictions.

To this end, therefore, we define a second nominal subsystem model to use for predictions in the ancillary controller:
\begin{equation}
\hat{x}_i^+=A_{ii}\hat{x}_i+B_{ii}\hat{u}_i+\bar{w}_i.
\label{eqn:nom_sys1}
\end{equation}
The disturbance term $\bar{w}_i$ is composed of the predictions
$(\bar{x}_j,\bar{u}_j)$ gathered from each of the neighbours,
$j\in\mc{N}_i$, of agent $i$ such that $\bar{w}_i = \sum_{\inN{j}{i}}
A_{ij}\bar{x}_j+B_{ij}\bar{u}_j$ and $\bar{\mb{w}}_i \triangleq
\{\bar{w}_i(0),\ldots,\bar{w}_i(N)\}$. From this model, we define a
nominal state error $\bar{e}_i \triangleq \hat{x}_i - \bar{x}_i$, and
control error $\bar{f}_i=\hat{u}_i-\bar{u}_i$, whose dynamics evolve
as
\begin{equation*}
\bar{e}_i^+=A_{ii}\bar{e}_i+B_{ii}\bar{f}_i + \bar{w}_i
  \end{equation*}
It is this model that is employed in the following, ancillary optimal
control problem, $\hat{\mbb{P}}_i(\bar{e}_i;\bar{\mb{w}}_{i})$:
\begin{equation}
\hat{V}_i^0 (\bar{e}_i;\bar{\mb{w}}_{i}) = \min_{\bar{\mb{f}}_i} \sum_{k = 0}^{H-1} \ell_i(\bar{e}_i(k),\bar{f}_i(k))
\label{eqn:mpc_inner}
\end{equation}
subject to, for $k=0,\ldots,H-1$,
\begin{subequations}
\begin{align}
  \bar{e}_i(0) & = \bar{e}_i,\\
  \bar{e}_i(k+1) &= A_{ii}\bar{e}_i(k)+ B_{ii}\bar{f}_i(k)+ \bar{w}_i(k), \label{eqn:cons_errdyn}\\
  \bar{e}_i(k) &\in\beta_i^x\mbb{X}_i, k = 0,\ldots,H-1\label{eqn:cons_erre}\\
	\bar{f}_i(k) &\in\beta_i^u\mbb{U}_i, k = 0,\ldots,H-1\label{eqn:cons_errf}\\
	\bar{e}_i(H) &= 0. \label{eqn:cons_errter}
\end{align}
\end{subequations}

In this problem, the decision variable is the sequence of controls
$\bar{\mb{f}}_i = \{\bar{f}_i(0),\dots,\bar{f}_i(H-1)\}$; the horizon
is $H$. The cost function is the same as in the main
problem. The parameter $\bar{\mb{w}}_{i}$ denotes the collection of
disturbance predictions for subsystems $j \in\mc{N}_i$. The state and
input constraints are, similar to in the main problem, scaled by
factors $\beta^x_i \in (0,1]$ and $\beta_i^u \in (0,1]$; detailed
design steps are given in Section~\ref{sec:constant_selec}.
  
  The solution of $\hat{\mbb{P}}_i(\bar{e}_i,\bar{\mb{w}}_i)$ defines
  an implicit control law
  \begin{equation*}
    f_i = \bar{f}_i^0(0;\bar{e}_i,\bar{\mathbf{w}}_i).
  \end{equation*}
  However, this alone is not sufficient to guarantee the recursive
  feasibility and stability properties that we seek. In particular, if $u_i = \bar{\kappa}_i(\bar{x}_i) + \bar{f}_i^0(0;\bar{e}_i,\bar{\mathbf{w}}_i)$ then
  \begin{equation*}
    e_i^+ - \bar{e}_i^+ = A_{ii}(e_i - \bar{e}_i) + (w_i - \bar{w}_i),
  \end{equation*}
  which is unsatisfactory because the error dynamics here depend on
  only the spectral radius of $A_{ii}$: if $A_{ii}$ is unstable, the
  mismatch between true error $e_i$ and planned error $\bar{e}_i$
  diverges. The next section describes how this problem is overcome by
  adding an extra feedback term to the ancillary control law.
    
  \subsection{Modified ancillary control law}

  We define $\hat{e}_i \triangleq e_i - \bar{e}_i$ and $\hat{w}_i
\triangleq w_i - \bar{w}_i$; by definition, $x_i = \bar{x}_i + e_i =
\bar{x}_i + \bar{e}_i + \hat{e}_i$, and thus we seek to regulate
$\bar{x}_i$, $\bar{e}_i$ and $\hat{e}_i$ to zero. To this end, we add
another control, $\hat{f}_i$, to the ancillary control law, \ie~$f_i =
\bar{f}_i^0(0;\bar{e}_i,\bar{\mb{w}}_i) + \hat{f}_i$, so that
\begin{equation*}
	\hat{e}_i^+=A_{ii}\hat{e}_i + B_{ii}\hat{f}_i+\hat{w}_i.
\end{equation*}
With an appropriate choice of
feedback law $\hat{f}_i = \mu_i(\hat{e}_i)$, then, this error can be
regulated and guaranteed to remain within an invariant set around the
origin, despite the disturbance $\hat{w}$. Therefore, the approach we
take to designing the additional feedback term is based on the concept
of \emph{robust control invariant} (RCI) sets~\citep{RKM+07} and their corresponding
invariance-inducing control laws.
\begin{defn}[RCI set]
A set $\mc{R}$ is \emph{robust control invariant} (RCI) for a system $x^+=f(x,u,w)$ and constraint set $\mbb{X}$, $\mbb{U}$ and $\mbb{W}$ if (i) $\mc{R}\subset\mbb{X}$ and (ii) for all $x\in\mc{R}$, there exists a $u\in\mbb{U}$ such that $x^+=f(x,u,w)\in\mc{R}$, $\forall w\in\mbb{W}$. 
\label{def:RCI_set}
\end{defn}
Given a RCI set $\mc{R}$, Definition~\ref{def:RCI_set} implies the
existence of a control law $\mu\colon\mbb{R}^m\mapsto\mbb{R}^n$, such
that the set mapping $\mu(\mc{R})\triangleq\{\mu(x):x \in \mc{R}\}
=\{u\in\mbb{U}: x^+\in\mc{R}, \forall w\in\mbb{W}\}$ is nonempty.
%
%% The set
%% $\mu(\mc{R})$ is the set of all the control actions that preserve the
%% invariance property.
%
Thus, given a RCI set, $\hat{\mc{R}}_i$, for the dynamics of the error
$\hat{e}_i$, the respective control action is chosen as $\hat{f}_i =
\mu_i(\hat{e}_i)$. The existence, design and computation of this invariant set
and control law is discussed in detail on
Section~\ref{sec:constant_selec}. For now, we note that the modified
ancillary control law is
\begin{equation*}
\hat{\kappa}_i(\bar{e}_i,\hat{e}_i,\bar{\mb{w}}_i) = \bar{f}^0_i(0;\bar{e}_i,\bar{\mb{w}}_{i}) + \mu_i(\hat{e}_i),
\end{equation*}
comprising the ancillary MPC control law plus the additional feedback term, and the overall control law for subsystem $i$ is
\begin{equation*}
u_i = \bar{\kappa}_i(\bar{x}_i) + \hat{\kappa}_i(\bar{e}_i,\hat{e}_i;\bar{\mb{w}}_{i}) = \bar{u}_i^0(0;\bar{x}_i) + \bar{f}^0_i(0;\bar{e}_i,\bar{\mb{w}}_{i}) + \mu_i(\hat{e}_i).
\end{equation*}
The structure of this three-term controller is worth remarking upon: the first term regulates the nominal state $\bar{x}_i$, while the second term regulates the planned error, accounting for planned (nominal) states and inputs of neighbours. The third term regulates the unplanned errors that arise from using nominal, rather than true, dynamics in the optimal control problems.

\subsection{Distributed Control Algorithm}\label{sec:dist_alg}
The optimization problems $\mathbb{P}_i(\bar{x}_i)$ and
$\hat{\mbb{P}}_i(\bar{e}_i,\bar{\mb{w}}_{i})$ are used in the
following algorithm.
\begin{alg}[NeDMPC for subsystem $i$]~

\textbf{Initial data}: Sets $\mbb{X}_i$, $\mbb{U}_i$; matrices
$(A_{ij},B_{ij})$ for $j \in \mc{N}_i$; constants $\alpha_i^x$,
$\alpha_i^u$, $\beta_i^x$,$\beta_i^u$; states $\bar{x}_i(0) = x_i(0)$, $\bar{e}_i = 0$, $\bar{\mb{w}}_i = 0$, $\hat{V}_i^* = +\infty$.

\textbf{Online Routine}:
\begin{enumerate}

\item\label{step:solve_outer} At time $k$, controller state $\bar{x}_i$, solve
  $\mathbb{P}_i(\bar{x}_i)$ to obtain $\bar{\mb{u}}_i^0$ and $\bar{\mb{x}}_i^0$.
  %% $\bar{u}_i^0 = \bar{\kappa}_i(\bar{x}_i)$ and $\bar{\mb{w}}^0_{ji}(\bar{x}_i) = \bigl((A_{ij}\bar{\mb{x}}_i^0(\bar{x}_i) +B_{ji}\bar{\mb{u}}_i^0(\bar{x}_i)\bigr)$, $\forall j\in\mc{N}_i$.
%
\item\label{step:transmit} Transmit $(\bar{\mb{x}}_i^0,\bar{\mb{u}}_i^0)$ to controllers $j \in \mc{N}_i$.
\item Compute $\bar{\mb{w}}_i^0 = \{\bar{{w}}^0_i(l)\}_l$ from received $(\bar{\mb{x}}_j^0,\bar{\mb{u}}_j^0)$, where $\bar{{w}}^0_i(l) = \sum_{j \in \mc{N}_i} (A_{ij}\bar{x}^0_j(l) + B_{ij}\bar{u}^0_j(l))$, $l = 0\dots N$.
\item\label{step:solve_inner} At controller state $\bar{e}_i$, solve
  $\hat{\mbb{P}}_i(\bar{e}_i;\bar{\mb{w}}^0_{i})$ to obtain
  $\bar{f}_i^0$: if feasible and $\hat{V}^0_i(\bar{e}_i;\bar{\mb{w}}^0_{i}) \leq \hat{V}_i^*$, set
  $\bar{\mb{w}}_{i} = \bar{\mb{w}}^0_{i}$ and $\hat{V}^*_i = \hat{V}_i^0(\bar{e}_i;\bar{\mb{w}}^0_i)$; else, solve
    $\hat{\mbb{P}}_i(\bar{e}_i;\bar{\mb{w}}_{i})$ for $\bar{f}_i^0$.
\item\label{step:law} Measure plant state $x_i$, calculate $\hat{e}_i = x_i - \bar{x}_i - \bar{e}_i$, and apply $u_i = \bar{u}_i^0 + \bar{f}_i^0 + \mu_i(\hat{e}_i)$.
\item Update controller states as $\bar{x}_i^+ = A_{ii}\bar{x}_i+B_{ii}\bar{u}^0_i$ and $\bar{e}_i^+ = A_{ii}\bar{e}_i+B_{ii}\bar{f}_i^0+\bar{w}_i$---where $\bar{w}_i=\bar{w}_i(0)$---$\bar{\mb{w}}^+_i = \{\bar{w}_i(1),\dots,\bar{w}_i(N),0\}$, and $V^{*+}_i = V_i^* - \ell_i(\bar{e}_i,\bar{f}^0_i)$.

\item Set $k=k+1$, $\bar{x}_i = \bar{x}_i^+$, $\bar{e}_i = \bar{e}_i^+$, $\bar{\mb{w}}_i = \bar{\mb{w}}_i^+$, $V^*_i = V_i^{*+}$, and go to Step~\ref{step:solve_outer}.

\end{enumerate}
\label{alg:nedmpc}
\end{alg}

In step~\ref{step:solve_inner}, the ancillary problem is solved using
the new disturbance sequence, $\bar{\mb{w}}_i^0$, formed from the state
and input sequences of other subsystems just optimized in
Step~\ref{step:solve_outer}. If this problem is infeasible, or the optimal cost does not decrease sufficiently with respect to the previous solution, the
problem is re-solved albeit with the previous disturbance sequence,
$\bar{\mb{w}}_i$; as will be shown, this problem remains feasible even
when the new problem is not, and in fact a feasible solution can be
generated without solving the problem.

\begin{rem}
  Guaranteeing the recursive feasibility of the ancillary problem is
  simple when the disturbance sequence is unchanging, but when the
  latter changes it is a non-trivial challenge. On the other hand, the
  feasibility of the ancillary problem depends on the horizon $H$,
  and---in view of the fact that $\bar{\mb{w}}_i$ is a sequence of $N$
  disturbances, with $\bar{w}_i(N) = 0$ always---it is suggested that
  $H \geq N+1$. In that case, $\bar{w}_i(k) = 0$ for prediction step $k \geq N$.
  \end{rem}

  This completes the description of the approach, including control
  problems and the algorithm. In Section~\ref{sec:constant_selec}, we
  present a comprehensive approach to designing the
  invariance-inducing controller $\mu_i$ and the set scaling
  parameters $a_i^x$, $a_i^u$, $\beta_i^x$ and $\beta_i^u$. Before
  that, we establish recursive feasibility and stability of the
  approach, which points to necessary and sufficient conditions on the
  scaling parameters that are useful later in developing the controller design
  process.

\section{Recursive feasibility and stability}\label{sec:stab_feas}
Recursive feasibility is the main challenge for this
approach. In contrast to conventional tube MPC, which uses linearity
of the error dynamics and robust positive invariant (RPI) sets to
allow the exact determination of constraint tightening margins for
robustness, here the error dynamics are nonlinear and the constraint
tightening is via scaling factors. In this section, we aim to
establish conditions under which the proposed control scheme is
recursively feasible and stable. Our approach here uses the notion of robust control invariant (RCI)
sets~\citep{RKM+07}: we show that, by suitable choices of scaling
factors $\alpha_i^x$, $\alpha_i^u$, $\beta_i^x$ and $\beta_i^u$, the error states
of the controlled system evolve within bounded RCI sets, which may be
used to guarantee constraint satisfaction and feasibility; however, we
do not seek to obtain an explicit representation of the RCI set, but
merely rely on its existence---an implicit form of invariance.

In order to establish robust constraint satisfaction, it is sufficient
to show that the state $x_i$ of subsystem $i$ is contained within a
set, say $\mc{X}_i$, that is robust positively invariant for the
dynamics $x_i^+ = A_{ii}x_i + B_{ii}u_i + w_i$ and constraint sets
$(\mbb{X}_i,\mbb{U}_i,\mbb{W}_i)$ under the control law
$u_i = \kappa_i(x_i)$: that is, given
$x_i \in \mc{X}_i \subseteq \mbb{X}_i$,
$A_{ii}x_i + B_{ii}\kappa_i(x_i) + w_i \in \mc{X}_i$ and
$\kappa_i(x_i) \in \mbb{U}_i$. In our approach, however, the true
state satisfies
$x_i = \bar{x}_i + e_i = \bar{x}_i + \bar{e}_i + \hat{e}_i$, about
which the following is known: the nominal state $\bar{x}_i$ resides
within $\bar{\mc{X}}_i^N$, defined as the feasibility region of
$\mbb{P}_i(\bar{x}_i)$:
\begin{equation}
  \bar{\mc{X}}_i^N\triangleq  \{\bar{x}_i : \bar{\mc{U}}_i^N(\bar{x}_i)\neq\emptyset\},\label{eq:outer_region}
  \end{equation}
where $\bar{\mc{U}}_i^N(\bar{x}_i)\triangleq\{\bar{\mb{u}}_i : \ts{\eqref{eqn:cons_init}--\eqref{eqn:cons_term} are satisfied}\}$; the planned error $\bar{e}_i$, given $\bar{\mb{w}}_i$, resides within $\bar{\mc{E}}_i^N(\bar{\mb{w}}_i)$, the feasibility region of $\hat{\mbb{P}}_i(\bar{e}_i;\bar{\mb{w}}_i)$:
\begin{equation}
\bar{\mc{E}}^N_i(\bar{\mb{w}}_{i}) \triangleq  \{\bar{e}_i\in\mbb{X}_i: \bar{\mc{F}}_i^N(\bar{e}_i;\bar{\mb{w}}_{i})\neq\emptyset\},\label{eq:inner_region}	
\end{equation}
where $\bar{\mc{F}}_i^N(\bar{e}_i;\bar{\mb{w}}_{i}) \triangleq
\{\mb{f}_i: \ts{\eqref{eqn:cons_errdyn}--\eqref{eqn:cons_errter} are
  satisfied}\}$; finally, we suppose that the unplanned error
$\hat{e}_i$ resides within some set $\hat{\mc{R}}_i$.  Then our task
is to develop conditions under which $x_i \in \bar{\mc{X}}_i^N \oplus \bar{\mc{E}}_i^N(\bar{\mb{w}}_i) \oplus \hat{\mc{R}}_i$ implies (i)
$x_i^+ \in \bar{\mc{X}}_i^N \oplus \bar{\mc{E}}_i^N(\bar{\mb{w}}^+_i)
\oplus \hat{\mc{R}}_i$, (ii) all constraints are satisfied, and (iii)
all MPC problems remain feasible (\ie~$\bar{x}_i^+ \in
\bar{\mc{X}}_i^N$ and $\bar{e}_i^+ \in
\bar{\mc{E}}_i^N(\bar{\mb{w}}^+_i)$). To this end, noting that $\bar{\mc{X}}_i^N \subseteq \alpha_i^x \mbb{X}_i$ by construction, we make the following assumptions, which may also be interpreted as design conditions that guide Section~\ref{sec:constant_selec}:

%% In
%% Section~\ref{sec:constant_selec}, we link the existence of suitable
%% constraint scaling factors to the feasibility of an LP problem;

%% The solution of the ancillary problem
%% $\hat{\mbb{P}}_i(e_i;\bar{\mb{z}}_{-i})$ yields the control law, $f_i
%% = \hat{\kappa}_i(\bar{e}_i,\hat{e}_i,\bar{\mb{w}}_{i})
%% =\bar{f}^0_i(0;\bar{e}_i,\bar{\mb{w}}_{i})+ \hat{f}_i(\hat{e}_i)$. Our
%% aim is to prove that this law is invariant for the error $e_i$, by
%% exploiting the structure of the disturbance sets.

%% The true state of each subsystem $i$ satisfies $x_i=\bar{x}_i+e_i$, and the error $e_i$ can be expressed in terms of the other errors since $e_i = \bar{e}_i + \hat{e}_i$. Given an  RCI set $\hat{\mc{R}}_i$, following definition~\ref{def:RCI_set}, for the dynamics of the error $\hat{e}_i$, and using proposition 1 from \cite{MSR05}, if $e_i\in\bar{e}_i\oplus\hat{\mc{R}}_{i}$ then $e_i^+\in\bar{e}_i^+\oplus\hat{\mc{R}}_{i}$. Consequently, it follows that $x_i\in\bar{x}_i\oplus\bar{e}_i\oplus\hat{\mc{R}}_i$, and $x_i^+\in\bar{x}_i^+\oplus\bar{e}_i^+\oplus\hat{\mc{R}}_i$. On the other hand, the domains of the value functions for the optimisation problems $\mbb{P}_i(\bar{x}_i)$ and $\hat{\mbb{P}}_i(\bar{e}_i,\bar{\mb{w}}_{i})$ are respectively,
%% %

%% %
%% Where the  and . These sets satisfy the following assumptions
%

\begin{assum}
\label{assump:invariance_R_h}
The set $\hat{\mc{R}}_i$ is RCI for the system $\hat{e}_i^+ =
A_{ii}\hat{e}_i + B_{ii}\hat{f}_i + \hat{w}_i$ and constraint set
$(\xi_i^x\mbb{X}_i, \xi_i^u\mbb{U}_i, \hat{\mbb{W}}_i)$, for some
$\xi_i^x\in [0,1)$ and $\xi_i^u\in [0,1)$, and where $\hat{\mbb{W}}_i \triangleq \bigoplus_{j \in \mc{N}_i} (1-\alpha^x_j)A_{ij}\mbb{X}_j \oplus (1-\alpha^u_j)B_{ij}\mbb{U}_j$. An invariance inducing control law for $\hat{\mc{R}}_i$ is $\hat{f}_i = \mu_i(\hat{e}_i)$.
 \end{assum}

% \begin{assum}
% \label{assump:contr_E}
% 	The feasibility sets of the ancillary problem satisfy
%         $\mc{E}_i^N(\bar{\mb{w}}_{i})\subset\beta_i^x\mbb{X}_i$ for
%         the selected $\beta_i^x\in [0,1)$ and are non-empty all $\bar{\mb{w}}_{i}
%           \in \bar{\mc{W}}^N_i$, where $\bar{\mc{W}}_i^N \triangleq
%           \bar{\mbb{W}}_i \times \bar{\mbb{W}}_i \times \dots \times
%           \bar{\mbb{W}}_i \times \{0\}$, and $\bar{\mbb{W}}_i =
%           \bigoplus_{j \in \mc{N}_i} (\alpha_j^x A_{ij} \mbb{X}_j
%           \oplus \alpha_j^u B_{ij} \mbb{U}_j)$.
% \end{assum}

\begin{assum}
  \label{assump:constraint_sat}
  The constants $(\alpha_i^x,\beta_i^x,\xi_i^x)$ and
  $(\alpha_i^u,\beta_i^u,\xi_i^u)$ are chosen such
  $\alpha_i^x+\beta_i^x+\xi_i^x\leq 1$ and
  $\alpha_i^u+\beta_i^u+\xi_i^u\leq 1$.
  \end{assum}

  The following result establishes recursive feasibility and
  constraint satisfaction under these assumptions. To aid the
  statement of the result, we first make the following definitions:
  $\bar{\mbb{W}}_i = \bigoplus_{j \in \mc{N}_i} (\alpha_j^x A_{ij}
  \mbb{X}_j \oplus \alpha_j^u B_{ij} \mbb{U}_j)$ is the set of
  admissible disturbances arising from the solutions of the main
  optimal control problems for $j \in \mc{N}_i$;
  $\bar{\mc{W}}^N_i \triangleq \bar{\mbb{W}}_i \times \bar{\mbb{W}}_i
  \times \dots \times \bar{\mbb{W}}_i \times \{0\}$ is the sequence of
  such sets. Given a disturbance sequence
  $\bar{\mb{w}}_i = \{\bar{w}_i(0),\dots,\bar{w}_i(N-1),0\} \in
  \bar{\mc{W}}^N_i$,
  $\bar{\mb{w}}^+_i = \{\bar{w}_i(1),\dots,\bar{w}_i(N-1),0,0\}$ is
  the tail of that sequence.
  \begin{prop}[Recursive feasibility]
    Suppose that Assumptions~\ref{assump:invariance_R_h}--\ref{assump:constraint_sat} hold. Then, for each subsystem $i \in \mc{M}$,
\begin{enumerate}[(i)]
\item If $\bar{x}_i \in \bar{\mc{X}}_i^N$ then $\bar{x}^+_i \in \bar{\mc{X}}_i^N$.
\item If $\bar{e}_i \in \bar{\mc{E}}_i^N(\bar{\mb{w}}_i)$, for some
  $\bar{\mb{w}}_i \in \bar{\mc{W}}^N_i$, then
  $\bar{e}^+_i \in \bar{\mc{E}}_i^N(\bar{\mb{w}}^+_i)$.
\item Given $\bar{x}_i(0)=x_i(0)\in \bar{\mc{X}}^N_i$, the subsystem
  $x_i^+=A_{ii}x_i+B_{ii}u_i+w_i$ under the control law
  $u_i =\bar{\kappa}_i(\bar{x}_i) +
  \hat{\kappa}_i(\bar{e}_i,\hat{e}_i;\bar{\mb{w}}_{i}) =
  \bar{u}_i^0(0;\bar{x}_i) + \bar{f}^0_i(0;\bar{e}_i,\bar{\mb{w}}_{i})
  + \mu_i(\hat{e}_i)$ satisfies $x_i\in\mbb{X}_i$ and
  $u\in\mbb{U}_i$ for all time.
\end{enumerate}
\end{prop}
\ifproves
\begin{pf}
  For part (i), because the nominal model is linear,
  $\alpha^x_i\mbb{X}_i$ and $\alpha^u_i\mbb{U}_i$ are PC-sets, and the
  terminal constraint is control invariant, the set $\bar{\mc{X}}_i^N$
  is compact, contains the origin and satisfies
  $\bar{\mc{X}}_i^N \supseteq \bar{\mc{X}}_i^{N-1} \supseteq \dots
  \supseteq \bar{\mc{X}}_i^0 = \{0\}$. Moreover, $\bar{\mc{X}}_i^N$ is
  positively invariant for
  $\bar{x}_i^+ = A_{ii}\bar{x}_i + B_{ii}\bar{\kappa}_i(\bar{x}_i)$, which
  is sufficient to prove the claim. (For a detailed proof,
  see~\citet[Proposition 2.11]{RM_mpc_book}.) The same arguments applied to
  $\bar{\mc{E}}_i(\bar{\mb{w}})$ establish part (ii).

  For (iii), suppose that at time $k$, $\bar{x}_i \in \bar{\mc{X}}^N_i$,
  $\bar{e}_i \in \bar{\mc{E}}_i^N(\bar{\mb{w}}_i)$ with
  $\bar{\mb{w}}_i \in \bar{\mc{W}}^N_i$, and
  $\hat{e}_i \in \hat{\mc{R}}_i$. Then
  ${x}_i \in \bar{\mc{X}}^N_i \oplus \bar{\mc{E}}_i^N(\bar{\mb{w}}_i)
  \oplus \hat{\mc{R}}_i \subseteq \alpha^x_i \mbb{X}_i \oplus
  \beta^x_i \mbb{X}_i \oplus \xi^x_i \mbb{X}_i =
  (\alpha^x_i+\beta^x_i+\xi^x_i)\mbb{X}_i \subseteq \mbb{X}_i$. The
  applied control is
  $u_i = \bar{u}_i^0(0;\bar{x}_i) +
  \bar{f}_i^0(0;\bar{e}_i,\bar{\mb{w}}_i) + \mu_i(\hat{e}_i) \in
  \alpha^u_i\mbb{U}_i\oplus\beta^u_i\mbb{U}\oplus\xi^u_i\mbb{U}_i
  \subseteq \mbb{U}_i$. Then, because of parts (i) and (ii),
  ${x}_i^+ = A_{ii}x_i + B_{ii}u_i + w_i \in \bar{\mc{X}}^N_i \oplus
  \bar{\mc{E}}_i^N(\bar{\mb{w}}_i^+) \oplus \hat{\mc{R}}_i$. To
  complete the proof, however, we must consider the possibility that
  the disturbance sequence at the successor state is
  $\bar{\mb{w}}_i^0 \neq \bar{\mb{w}}_i^+$: in that case, if
  $\hat{\mbb{P}}_i(\bar{e}_i^+;\bar{\mb{w}}_i^0)$ is feasible then
  ${x}_i^+\in \bar{\mc{X}}^N_i \oplus \bar{\mc{E}}_i^N(\bar{\mb{w}}_i^0)
  \oplus \hat{\mc{R}}_i$, which is still within $\mbb{X}_i$ by
  construction, and
  $u_i = \bar{u}_i^0(0;\bar{x}^+_i) +
  \bar{f}_i^0(0;\bar{e}^+_i,\bar{\mb{w}}^0_i) + \mu_i(\hat{e}^+_i)
  \subseteq \mbb{U}_i$. If
  $\hat{\mbb{P}}_i(\bar{e}_i^+;\bar{\mb{w}}_i^0)$ is not feasible,
  then $\hat{\mbb{P}}_i(\bar{e}_i^+;\bar{\mb{w}}_i^+)$ \emph{is}
  feasible (by the tail), and
  $u_i = \bar{u}_i^0(0;\bar{x}^+_i) +
  \bar{f}_i^0(0;\bar{e}^+_i,\bar{\mb{w}}^+_i) + \mu_i(\hat{e}^+_i)
  \subseteq \mbb{U}_i$. This establishes recursive feasibility of the
  algorithm.

  Finally, if, at time $0$, $\bar{x}_i = x_i \in \bar{\mc{X}}_i^N$ then
  $\bar{{e}}_i = 0$. Moreover, if $\bar{\mb{w}}_i = 0$,
  then---trivially---$\bar{{e}}_i \in \bar{\mc{E}}^N_i(0)$ and both
  the main and ancillary problems are feasible. By recursion,
  feasibility is retained at the next step, and the proof is complete.
  \qed
\end{pf}
\fi
Having established recursive feasibility and constraint satisfaction,
the main result follows. The following assumption is supposed to hold.

\begin{assum}[Decentralized stabilizability]
  The RCI control laws $u_i = \mu_i(x_i)$ asymptotically stabilize the
  system $x^+ = Ax+Bu$.
\end{assum}

\begin{thm}[Asymptotic stability]
For each $i \in\mc{M}$, (i) the origin is asymptotically stable for the composite subsystem
\begin{align*}
	\bar{x}_i^+& = A_{ii}\bar{x}_i + B_{ii}\bar{\kappa}_i(\bar{x}_i)\\
	\bar{e}_i^+& = A_{ii}\bar{e}_i + B_{ii}\bar{f}_i(0;\bar{e}_i,\bar{\mb{w}}_i) + \bar{w}_i.
\end{align*}
(ii) The origin is asymptotically stable for $x_i^+ = A_{ii}x_i + B_{ii}\kappa_i(x_i) + w_i$. The region of attraction is $\bar{\mc{X}}^N_i \subseteq \alpha^x_i\mbb{X}_i$.
\end{thm}

\ifproves
\begin{pf}
  For (i), asymptotic stability of $0$ for
  $\bar{x}_i^+ = A_{ii}\bar{x}_i + B_{ii}\bar{\kappa}_i(\bar{x}_i)$
  follows from the following facts: the value function
  $\bar{V}^0_i(\bar{x}_i)$ satisfies, for all $\bar{x}_i \in \bar{\mc{X}}^N_i$,
\begin{align*} 
\bar{V}^0_i(\bar{x}_i) &\geq \ell_i(\bar{x}_i,\bar{\kappa}_i(\bar{x}_i)),\\
\bar{V}^0_i(0) &= 0,\\
\bar{V}^0_i(\bar{x}^+_i) - \bar{V}^0_i(\bar{x}_i) &\leq - \ell_i(\bar{x}_i,\bar{\kappa}_i(\bar{x}_i)). 
\end{align*}
Therefore $\{\bar{V}^0_i(\bar{x}_i)\} \to 0$ and $\bar{x}_i \to 0$, $\bar{u}_i \to 0$. Similar arguments applied to $\hat{V}_i^0(\bar{e}_i;\bar{\mb{w}}_i)$---together with the fact that because $\bar{w}_i$ is a linear function of $(\bar{x}_j,\bar{u}_j)$ for $j \in \mc{N}_i$, then $\bar{w}_i \to 0$ and $\bar{\mb{w}}_i \to 0$---establish that $\bar{e}_i \to 0$; the possibility that $\hat{V}_i^0(\bar{e}_i;\bar{\mb{w}}_i)$ does not attain the necessary decrease between $(\bar{e}_i,\bar{\mb{w}}_i)$ and $(\bar{e}^+_i,\bar{\mb{w}}_i^0)$ (where $\bar{\mb{w}}_i^0 \neq {\mb{w}}_i)$ is eliminated by the checking step in the algorithm.

For (ii), because $x_i \in \bar{x}_i + \bar{e}_i + \hat{e}_i$ and $\bar{x}_i, \bar{e}_i \to 0$, then $x_i \to \hat{e}_i$ and $u_i \to \mu_i(x_i)$. Under the decentralized stabilizability assumption, then $x \to 0$ and so each $x_i \to 0$.\qed
\end{pf}
\fi
\section{Selection of the scaling constants}\label{sec:constant_selec}
In this section, a methodology is given for the design of the scaling constants $\alpha_i^x$, $\alpha_i^u$, $\beta_i^x$ and $\beta_i^u$ for the main and ancillary problems, and the RCI controller $\mu_i(\cdot)$. The approach we take is to employ the optimized RCI set design proposed by~\citet{RKM+07}; however, we do not explicitly construct the set $\hat{\mc{R}}_i$, but use the optimization to produce the scaling constants and the control law.
\subsection{Revision of optimized robust control invariance}

In~\cite{RKM+07}, the problem of computing an RCI set for $x^+=Ax+Bu+w$
and $(\mbb{X},\mbb{U},\mbb{W})$, with $\mbb{W}$ a C-set, is posed as
linear programming (LP) problem. The set, and corresponding control
set, are the polytopes
\begin{align*}
\mc{R}_h(\mb{M}_h) = \bigoplus\limits_{l=0}^{h-1}D_l(\mb{M}_h)\mbb{W}, && \mu(\mc{R}_h(\mb{M}_h)) = \bigoplus\limits_{l=0}^{h-1} M_l \mbb{W},
\label{eqn:rci_set}
\end{align*}
where the matrices $D_l(\mb{M}_h), l = 0\dots h$ are defined as
\begin{align*}
  D_0(\mb{M}_h) = I, &&
  D_l(\mb{M}_k) \triangleq A^l + \sum_{j=0}^{l-1} A^{l-1-j}BM_j, l \geq 1
  \end{align*}
with $M_j \in \mbb{R}^{m\times n}$ and $\mb{M}_h \triangleq
(M_0,M_1,\dots,M_{h-1})$, such that $D_h(\mb{M}_h) = 0, h \geq n$. The
set of matrices that satisfy these conditions is given by
$\mbb{M}_h\triangleq \{\mb{M}_h : D_h(\mb{M}_h)=0\}$.  Constraint
satisfaction is guaranteed if $\mc{R}_h(\mb{M}_h) \subseteq \eta
\mbb{X}$ and $\mu(\mc{R}_h(\mb{M}_h)) \subseteq \theta \mbb{U}$, with
$(\eta,\theta) \in [0,1] \times [0,1]$.

The optimization problem defined to compute these sets is
\begin{equation*}
\mbb{P}^\mc{R}_h: \min \{ \delta : \gamma \in \Gamma\},
\end{equation*}
where $\gamma=(\mb{M}_h,\eta,\theta,\delta)$, and the set
$\Gamma=\{\gamma : \mb{M}_h \in\mbb{M}_h,
\mc{R}_h(\mb{M}_h)\subseteq\eta\mbb{X},
\mu(\mc{R}_h(\mb{M}_h))\subseteq\theta\mbb{U},
(\eta,\theta)\in[0,1]\times
   [0,1],\ q_{\eta}\eta+q_{\theta}\theta\leq\delta\}$; $q_{\eta}$
   and $q_{\theta}$ are weights to express a preference for the
   relative contraction of state and input constraint
   sets. Feasibility of this problem is linked to the existence of an
   RCI set: if $\mbb{P}^\mc{R}_h$ is feasible, then
   $\mc{R}_h(\mb{M}_h)$ satisfies the RCI properties~\citep{RKM+07}.

   \subsection{Design procedure for each subsystem}

   Recall that in the control algorithm proposed in the previous
   section, the state error $e_i = x_i - \bar{x}_i$ was decomposed
   into planned error $\bar{e}_i = \hat{x}_i - \bar{x}_i$ and an
   unplanned error $\hat{e}_i = x_i - \hat{x}_i$; thus,
   $e_i = \bar{e}_i+\hat{e}_i$.  Our aim is to determine the RCI
   control law $\hat{f}_i = \mu_i(\hat{e}_i)$ associated with the
   unplanned error dynamics
   $\hat{e}_i^+ = A_{ii}\hat{e}_i + B_{ii}\hat{f}_i + \hat{w}_i$. The
   principal challenge here is that it is not possible, \emph{a
     priori}, to define the unplanned error set
   $\hat{\mbb{W}}_i$. Instead, we consider that an RCI problem
   $\mbb{P}^\mc{R}_h$ is associated with the error dynamics
   $e_i^+ = A_{ii}e_i + B_{ii}f_i + w_i$ and constraint sets
   $(\mbb{X}_i,\mbb{U}_i,\mbb{W}_i)$, and call this problem
   $\mbb{P}^{\mc{R}_i}_h$, with the set $\mc{R}_{i,h}$ defined by
   adding appropriate $i$ subscripts to its generating sets and
   matrices. The rationale for this is as follows.

   The disturbance $w_i=\sum_{j\in\mc{N}_i}A_{ij}x_j + B_{ij}u_j$
   arising from the state and input coupling, is decomposed into two
   terms: $w_i = \bar{w}_i+\hat{w}_i$. The first term,
   $\bar{w}_i = \sum_{j\in\mc{N}_i}A_{ij}\bar{x}_j + B_{ij}\bar{u}_j$,
   is the planned disturbance obtained from the predictions, while the
   second term,
   $\hat{w}_i = \sum_{j\in\mc{N}_i}A_{ij}(x_j-\bar{x}_j) +
   B_{ij}(u_j-\bar{u}_j)$, is the unplanned disturbance. Since $\bar{x}_j \in \alpha_j^x\mbb{X}_j$, $\bar{u}_j \in \alpha_j^u\mbb{U}_j$ then $\bar{\mbb{W}}_i = \bigoplus_{j \in \mc{N}_i} (\alpha^x_i
   A_{ij}\mbb{X}_j \oplus \alpha^u_i B_{ij}\mbb{U}_j)$, In addition, if we bound $e_j\in(1-\alpha_j^x)\mbb{X}_j$ and $f_j\in(1-\alpha_j^u)\mbb{U}_j$, it is possible to write $\hat{\mbb{W}}_i = \bigoplus_{j \in \mc{N}_i} ((1-\alpha^x_i)
   A_{ij}\mbb{X}_j \oplus (1-\alpha^u_i) B_{ij}\mbb{U}_j)$ and so, $w_i \in \mbb{W}_i = \bar{\mbb{W}}_i \oplus \hat{\mbb{W}}_i$,\ie~$\bar{\mbb{W}}_i$ and $\hat{\mbb{W}}_i$ are summands of the known $\mbb{W}_i$. The next results follow directly from the definition of RCI sets and the results of~\cite{RKM+07}:
\begin{prop}
\label{prop:rci_ex}
Suppose Assumptions~\ref{assump:contr}--\ref{assump:disturbance}
hold. If $\mbb{P}^{\mc{R}_i}_h$ is feasible for $i \in \mc{M}$, then
$\mc{R}_{i,h}(\mb{M}_{i,h})$ is an RCI set for $e^+_i = A_{ii}e_i
+ B_{ii}f_i + w_i$ and $(\mbb{X}_i,\mbb{U}_i,\mbb{W}_i)$.
\end{prop}
\begin{prop}
\label{prop:rci_sub}
Suppose $\tilde{\mbb{W}}_i\subset\mbb{W}_i$ is a PC-set and a summand of
$\mbb{W}_i$. If $\mc{R}_{i,h}(\mb{M}_{i,h})$ is an RCI set for
$e^+_i = A_{ii}e_i + B_{ii}f_i + w_i$ and
$(\mbb{X}_i,\mbb{U}_i,\mbb{W}_i)$, then $\tilde{\mc{R}}_{i,h}(\mb{M}_{i,h}) = \bigoplus_{l=0}^{h-1}D_l(\mb{M}_h)\tilde{\mbb{W}}_i \subset \mc{R}_{i,h}(\mb{M}_{i,h})$ is an RCI set for $e^+_i = A_{ii}e_i + B_{ii}f_i + w_i$ and
$(\mbb{X}_i,\mbb{U}_i,\tilde{\mbb{W}}_i)$.
\end{prop}
The implication of the second result is that it is possible to first
determine an RCI set for the known disturbance set $\mbb{W}_i$, and
then, from that, determine an RCI set (with the same structure) for
the set $\hat{\mbb{W}}_i$, because the latter is a summand. Therefore,
the design is summarized as follows:
\begin{enumerate}
\item The problem $\mbb{P}^{\mc{R}_i}_h$ associated with the known $\mbb{W}_i$
  is solved to yield
  $\gamma_{i,h} = (\mb{M}_{i,h},\eta_i,\theta_i,\delta_i)$, where
  $\eta_i$ and $\theta_i$ are scalings of $\mbb{X}_i$ and $\mbb{U}_i$
  such that $\mc{R}_{i,h} \subset \eta_i\mbb{X}_i$ and
  $\mu(\mc{R}_{i,h}) \subset \theta_i \mbb{U}_i$ respectively.
\item Given that, under the RCI control law,
  $e_i \in \mc{R}_{i,h} \subset \eta_i\mbb{X}_i$ and
  $f_i \in \mu(\mc{R}_{i,h}) \subset \theta_i \mbb{U}_i$, we select
\begin{subequations}
\begin{align*}
	\alpha_i^x &= 1 - \eta_i\\
	\alpha_i^u &= 1 - \theta_i.	
\end{align*}
\label{eq:alpha}%
\end{subequations}
Then $x_i = \bar{x}_i + e_i \in \alpha^x_i\mbb{X}_i \oplus \eta_i\mbb{X}_i = \mbb{X}_i$, as required, with a similar expression for $u_i$.
\item The selection of suitable $\xi^x_i$ and
  $\xi^u_i$ is done by finding values such that the sets
  $\hat{\mc{R}}_{i,h}$ and $\mu_i(\mc{R}_{i,h})$ corresponding to the
  unplanned disturbance set
  $\hat{\mbb{W}}_i = \bigoplus_{j \in \mc{N}_i} ((1-\alpha^x_{j})
  A_{ij}\mbb{X}_j \oplus (1-\alpha^u_j)B_{ij}\mbb{U}_j)$ being
  contained within $\xi^x_i\mbb{X}_i$ and $\xi^u_i\mbb{U}_i$. The set
  $\hat{\mbb{W}}_i$ is computed and the RCI problem
  $\mbb{P}^{\hat{\mc{R}}_i}_h$ is solved for
  $\tilde{\gamma}_{(i,h)} =
  (\mb{M}_{i,h},\tilde{\eta}_i,\tilde{\theta}_i,\tilde{\delta}_i)$ to
  yield the scaling factors
  \begin{subequations}
\begin{align*}
	\xi_i^x &=  \tilde{\eta}_i\\
	\xi_i^u &=  \tilde{\theta}_i.	
\end{align*}
\label{eq:xi}%
\end{subequations}
\item The selection of the constants $\beta_i^x$ and $\beta^u_i$
  follows from Assumption~\ref{assump:constraint_sat} in order to
  satisfy constraint satisfaction
\begin{subequations}
\begin{align*}
	\beta_i^x &= 1 - \alpha^x_i - \xi^i_i\\
	\beta_i^u &= 1- \alpha^u_i - \xi^u_i.	
\end{align*}
\end{subequations}

\item The control law
  $\hat{f}_i = \mu_i(\hat{e}_i)$ is computed from the matrices $\mb{M}_{i,h}$, using  the minimal selection map procedure described in \citet{RKM+07}.

\end{enumerate}
\section{Illustrative example}\label{sec:sims}
To illustrate the feasibility of the design methodology, we consider
an example based on the system from \citet{FS12}, which comprises four
trucks, each with dynamics
\begin{equation*}
\dod{}{t}\begin{bmatrix}
r_i \\ v_i
\end{bmatrix}
=
A^c_{ii}
\begin{bmatrix}
r_i \\ v_i
\end{bmatrix}
+
\begin{bmatrix}  
0 \\ 100
\end{bmatrix}
u_i + w_i 
\end{equation*}
where $r_i$ is the displacement of truck $i$ from a datum, $v_i$ is
its velocity and $u_i$ is the control input (acceleration). The
disturbance $w_i$ arises via the coupling between trucks: truck 1
(mass $m_1=3$~kg) is coupled to truck 2 (mass $m_2=2$~kg) via a spring
(stiffness $k_{12}=0.5$) and damper ($h_{12}=0.2$). Likewise, truck 2
(mass $m_3=3$~kg) is coupled to truck 3 (mass $m_4=6$~kg) via
$k_{34}=1$ and $h_{34}=0.3$. Finally, truck 3 is coupled to truck 4
via $k_{23}=0.75$ and $h_{23}=0.25$. The initial conditions are
$x_1 = (1.8,-2)$, $x_2 = (0.5,5)$, $x_3 = (-0.9,-5)$, and
$x_4 = (-1.8,2)$. The problem is to steer the trucks to equilibrium
while satisfying constraints on displacement ($|r_i|\leq 2$), speed
($|v_i|\leq 8$) and acceleration ($|u_i|\leq 4$ for $i = 1,2,3$, and
$|u_4|\leq 6$). In each case, the controllers are designed with
$Q_{i} = I$, $R_i = 1$ and horizon $N=25$. Before applying Algorithm~\ref{alg:nedmpc} to the system
we obtain the scaling constants $(\alpha_i^x,\beta_i^x,\xi_i^x)$ and
$(\alpha_i^u,\beta_i^u,\xi_i^u)$ for each truck $i\in\mc{M}$---see
Table~\ref{table:constants} for the values obtained through the
procedure detailed in Section~\ref{sec:constant_selec}.
 \begin{table}[b]
   \centering\footnotesize
   \caption{Designed values of scaling factors.}
		\begin{tabular}{crrrr}
			\toprule
			  & Truck 1 & Truck 2 & Truck 3 & Truck 4\\
			\midrule
                  $\alpha_i^x$ & $0.9784$  &  $0.9228$  &  $0.9342$  &  $0.9816$ \\
		
			$\beta_i^x$ & $0.0199$  &  $0.0736$  &  $0.0628$  &  $0.0172$\\
		
			$\xi_i^x$ & $0.0017$  &  $0.0036$  &  $0.0029$  &  $0.0012$\\
		
			$\alpha_i^u$ & $0.9921$  &  $0.9808$  &  $0.9759$  &  $0.9910$\\
		
			$\beta_i^u$ & $0.0073$  &  $0.0183$  &  $0.0230$  &  $0.0084$\\
		
			$\xi_i^u$ & $0.0006$  &  $0.0009$  &  $0.0011$  &  $0.0006$\\
			\bottomrule			
	\end{tabular}	
	\label{table:constants}
\end{table}

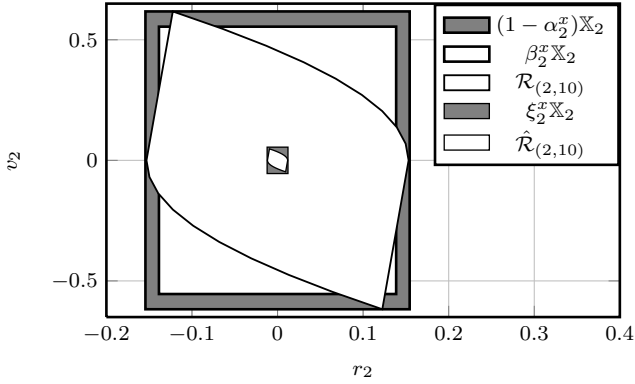
\begin{figure}[t]  
\centering\footnotesize
\begin{tikzpicture}

\begin{axis}[%
enlargelimits=false,
width=0.95\linewidth,
height=0.65\linewidth,
%scale only axis,
xmin=-0.2,
xmax=0.4,
xmajorgrids,
xlabel={$r_2$},
ymin=-0.65,
ymax=0.65,
ymajorgrids,
ylabel={$v_2$},
legend style={at={(1,1)},anchor=north east},
]

\addplot[area legend,solid,line width=1.0pt,draw=black,fill=gray,postaction={pattern=nwl,hatchspread=10pt,hatchshift=6pt}]
table[row sep=crcr] {%
x	y	z\\
0.154389268503421	-0.617557074013684	0\\
-0.154389268503421	-0.617557074013684	0\\
-0.154389268503421	0.617557074013684	0\\
0.154389268503421	0.617557074013684	0\\
}--cycle; \label{fig:s1}
\addlegendentry{$(1-\alpha_2^x)\mbb{X}_2$}
\addplot[area legend,solid,line width=1.0pt,draw=black,fill=white,postaction={pattern= nel,hatchspread=10pt,hatchshift=6pt}]
table[row sep=crcr] {%
x	y	z\\
0.138602083524619	-0.554408334098478	0\\
-0.138602083524619	-0.554408334098478	0\\
-0.138602083524619	0.554408334098478	0\\
0.138602083524619	0.554408334098478	0\\
}--cycle; \label{fig:s2}
\addlegendentry{$\beta^x_2\mbb{X}_2$}
\addplot[area legend,solid,line width=0.7pt,draw=black,fill=white,postaction={pattern= nwld}]
table[row sep=crcr] {%
x	y	z\\
0.122425035724894	-0.617557074013683	0\\
0.0642888953030425	-0.545501099695753	0\\
0.0132644787056307	-0.47525531213793	0\\
-0.0308090718775698	-0.406416658316838	0\\
-0.0680523163626325	-0.338581812628322	0\\
-0.0985453622287728	-0.271344619999059	0\\
-0.122327450632326	-0.204293462922138	0\\
-0.139396274658567	-0.137008534034273	0\\
-0.14970701930469	-0.0690589948366542	0\\
-0.153171110790664	0	0\\
-0.122425035724894	0.617557074013683	0\\
-0.0642888953030425	0.545501099695753	0\\
-0.0132644787056307	0.47525531213793	0\\
0.0308090718775698	0.406416658316838	0\\
0.0680523163626325	0.338581812628322	0\\
0.0985453622287728	0.271344619999059	0\\
0.122327450632326	0.204293462922138	0\\
0.139396274658567	0.137008534034273	0\\
0.14970701930469	0.0690589948366542	0\\
0.153171110790664	0	0\\
}--cycle; \label{fig:s3}
\addlegendentry{$\mc{R}_{(2,10)}$}

\addplot[area legend,solid,line width=0.5pt,draw=black,fill=gray]
table[row sep=crcr] {%
x	y	z\\
0.012187184978802	-0.054748739915206	0\\
-0.012187184978802	-0.054748739915206	0\\
-0.012187184978802	0.054748739915206	0\\
0.012187184978802	 0.054748739915206	0\\
}--cycle; \label{fig:s4}
\addlegendentry{$\xi^x_2\mbb{X}_2$}
\addplot[area legend,solid,line width=0.5pt,draw=black,fill=white,postaction={pattern=nel,hatchspread=1.6pt}]
table[row sep=crcr] {%
x	y	z\\
0.00945055585603578	-0.0476720924580428	0\\
0.00496275775936486	-0.0421097578749195	0\\
0.00102394658222076	-0.03668715999667	0\\
-0.00237829503522366	-0.0313731852925707	0\\
-0.00525327367159511	-0.0261366991901245	0\\
-0.00760717319445244	-0.0209463486969967	0\\
-0.00944302281050658	-0.0157703591502898	0\\
-0.010760644438319	-0.0105763236741387	0\\
-0.0115565786001393	-0.00533098384820625	0\\
-0.0118239878754135	0	0\\
-0.00945055585603578	0.0476720924580428	0\\
-0.00496275775936486	0.0421097578749195	0\\
-0.00102394658222076	0.03668715999667	0\\
0.00237829503522366	0.0313731852925707	0\\
0.00525327367159511	0.0261366991901245	0\\
0.00760717319445244	0.0209463486969967	0\\
0.00944302281050658	0.0157703591502898	0\\
0.010760644438319	0.0105763236741387	0\\
0.0115565786001393	0.00533098384820625	0\\
0.0118239878754135	0	0\\
}--cycle; \label{fig:s5}
\addlegendentry{$\hat{\mc{R}}_{(2,10)}$}
\end{axis}
\end{tikzpicture}%
%
\caption{For truck 2 (and $h=10$), the different scalings of the state
  constraint set $\mbb{X}_2$ and the RCI sets $\mc{R}_{2,10}$ and
  $\hat{\mc{R}}_{2,10}$: the main controller, ancillary controller and
  RCI controller operate within the regions $\alpha_i^x\mbb{X}_2$,
  $\beta_i^x\mbb{X}_2$ and $\eta_i^x\mbb{X}_2$ respectively; the space
  $(1-\alpha^x_2)\mbb{X}_2$ is divided between the ancillary
  controller ($\beta^x_2\mbb{X}_2$) and the RCI controller
  ($\xi_2^x\mbb{X}_2$) such that $1-\alpha_2^x = \beta_2^x + \xi_2^x$.}
    \label{fig:phase_tra}
\end{figure}

In Figure~\ref{fig:phase_tra}, the different scalings of the state
constraint sets are shown for truck 2, and also the corresponding RCI
sets. Thus, for truck 2, $92.28\%$ of the state constraint set is
allocated to the the main optimal control problem, which is concerned
with regulating the nominal subsystem (\ie~neglecting
interactions). On the other hand, the ancillary problem---which
regulates the planned errors---has $7.36\%$ of the original state
constraint sets. The remaining $0.36\%$ of the state constraint set is
allocated to the RCI control law to handle unplanned disturbances.

% \begin{table}[h!]
% 	\centering
% 		\begin{tabular}{|c|c|c|c|}
% 			\hline
% 			 & CMPC  &  DTMPC & Algorithm~\ref{alg:nedmpc}\\
% 			 \hline
% 			 Cost & 146.9547  & 151.4719  & 146.9547\\
% 			\hline
% 	\end{tabular}
% 	\caption{Values for the Cost function for different algorithms. }
% 	\label{table:tab:costs}
%       \end{table}

\section{Conclusions}\label{sec:conclusions}
A distributed MPC algorithm for dynamically coupled linear systems was
proposed. Subsystem controllers solve (once, at each time step) local
optimal control problems to determine control sequences and state
trajectories, and exchange information about these. The main feature
of the proposed algorithm is the use of a secondary MPC controller for
each subsystem, which acts on the shared plans of other subsystems and
aims to reject the uncertainty caused by neglecting interactions in
the main problems. Recursive feasibility and stability are guaranteed
under provided assumptions, and a design methodology was given for the
off-line selection of controller parameters and illustrated with an
example.

A key advantage of the proposed approach, in addition to the
guaranteed feasibility and stability and despite this being a
tube-based method, is the absence of invariant sets in the optimal
control problems. This makes the approach potentially applicable to
higher-dimensional subsystems.

\bibliography{extracted}             
\end{document}